\documentstyle[epsfig,11pt]{article}
\newcommand{\be}{\begin{equation}}
\newcommand{\ee}{\end{equation}}
\newcommand{\ben}{\begin{eqnarray}}
\newcommand{\een}{\end{eqnarray}}
\newcommand{\bc}{\begin{center}}
\newcommand{\ec}{\end{center}}

\begin{document}


\title{Accretion disc onto a static non-baryonic compact object}
\author{Diego F. Torres\thanks{dtorres@princeton.edu}
\\{\small
Physics Department, Princeton University, NJ 08544, USA} }

\date{\today }

\maketitle

\begin{abstract}
We study the emissivity properties of a geometrically thin,
optically thick, steady accretion disc about a static boson star.
Starting from a numerical computation of the metric potentials and
the rotational velocities of the particles in the vicinity of the
compact object, we obtain the power per unit area, the temperature
of the disc, and the spectrum of the emitted radiation. In order
to see if different central objects could be actually
distinguished, all these results are compared with the case of a
central Schwarzschild black hole of equal mass. We considered
different situations both for the boson star, assumed with and
without self-interactions, and the disc, whose internal
commencement can be closer to the center than in the black hole
case. We finally make some considerations about the Eddington
luminosity, which becomes radially dependent for a transparent
object. We found that, particularly at high energies, differences
in the emitted spectrum are notorious. Reasons for
that are discussed.\\

PACS Number(s): 04.40.Dg, 98.62.Mw, 04.70.-s

\end{abstract}



\section{Introduction}

\indent The possible existence of very massive non-baryonic
objects in the center of some galaxies is being studied since a
long time. As far as we are now aware, they were first
hypothesized by Tkachev \cite{TKA}, who also studied the
emissivity properties arising from particle anti-particle
annihilation processes. More recently, detailed studies of the
properties of neutrino ball scenarios have also been carried out
by Viollier and his collaborators \cite{NB}. In Ref.
\cite{GALAXY}, in addition, we have also explored whether
supermassive non-baryonic boson stars might be the central object
of some galaxies. To fix the situation to a particular case, we
have paid special attention to the Milky Way. This study had a
twofold aim. On one hand, it focused on what current dynamical
observational data have established regarding the properties of
the galactic center. We have concluded in this sense that scalar
stars fitted very well into these dynamical constraints. On the
other hand, we have also discussed what kind of observations could
actually distinguish between a supermassive black hole and a
boson star of equal mass, pointing out several possible tests.\\

In the case of our own Galaxy, recent observations probe the
gravitational potential at a radius larger than $\sim 10^{4}$
Schwarzschild radii \cite{Genzel}. The black hole scenario is the
current paradigm, but suggestions that the dark central objects
are indeed black holes are based only on indirect astrophysical
arguments, basically dynamical in nature, that could be sustained
by any small relativistic object other than a black hole, if it
exists \cite{kor95}. It is then advisable to explore possible
alternative scenarios. The final aim should be to devise
definitive tests that could observationally solve the issue. One
is to look for the event horizon projected onto the sky plane.
Although this is the key concept of both, Constellation-X
\cite{cons} and Maxim \cite{maxim} satellites, the needed spatial
resolution is still a dream for the future. It might be more
feasible to look for the realization of the recently introduced
concept of the shadow of a black hole \cite{shadow}. Although any
highly relativistic object would also produce a shadow, the
observed features might be enough
to decide whether an event horizon is present or not. \\

From the point of view of boson stars physics (see \cite{MIELKE}
for a recent review), we also need to know whether fundamental
scalars capable to form the stars do exist; observational tests in
this sense are highly desirable too. Only after the discovery of
the boson mass spectrum we shall be in position to determine which
galaxies, if any at all, could be modeled with such a center. In
recent works, Schunck and Liddle \cite{sch97}, Schunck and Torres
\cite{sch00}, and Capozziello et al. \cite{CAP00}, among others,
analyzed different observational effects that boson stars would
produce. Also, boson stars were proposed as sources for some of
the gamma ray bursts \cite{IWA}, and as a possible lens in a
gravitational lensing configuration \cite{dab00}, following recent
interest in analyzing the gravitational lensing phenomenon in
strong field regimes
\cite{vir98-vir00-tor98ab}. \\

However, as far as we know, literature lacks the study of an
accretion process, onto a boson star, either massive or
supermassive. Do the emissivity properties of the disc differ when
the central object, instead of being a black hole, is assumed to
be a boson star? Can we detect these differences? From where these
deviations, if any, come from? Can accretion onto boson stars help
model galactic centers? Of what kind? To completely answer these
questions would require the analysis of different models of
accretion discs, with various degrees of complexities. In this
first approach, we shall analyze the properties of the simplest,
steady, geometrically thin, and optically thick accretion disc
model, rotating onto a static boson star. We shall compare all our
results with those obtained using a black hole of
the same mass.\\

The fact that all circular orbits are stable for static stars (as
we shall show), can pose a problem to accretion scenarios upon
non-baryonic (with no surface) static objects. Accretion would
follow a series of stable circular orbits, loosing angular
momentum and radiating part of the generated heat. If particles
can always found a stable orbit, provided an enough amount of
time, they would all end up in the center, and should a way of
diverting them from there not exist, we would confront the
formation of a baryonic black hole in the center of every
non-baryonic star subject to overdense environments. This problem
have apparently (as far as we are aware \footnote{We acknowledge
e-mail discussions with Dr. Tkachev in early 2000 on issues
related to this point.}) been not clearly mentioned ever before in
the literature of boson star solutions (see \cite{GALAXY}),
although we have no other option than confront it if we are to
talk about any
astrophysically relevant use of these scalar star models.  \\

This problem can, however, be alleviated in more general
situations. In the rotating case, particles orbits were analyzed
by Ryan \cite{ryan} (see his section IV). He has shown that
circular geodesic orbits are not stable beyond a given point,
located at about 5/3 times the radius of the doughnut hole which
appears in rotating boson star solutions. He has considered the
swirling of a stellar size object (a black hole, or neutron star)
within a supermassive rotating boson star and studied the
gravitational wave emission as a mechanism for detection.
Accretion would not continue in circular orbits since there is
none after that point. Also, we are just analyzing the case in
which a particle continues to travel in geodesics within the star
interior, disregarding any possible influence of the boson star
matter. And there is also the fact that two body encounters will
be unavoidable in the innermost regions of the boson star, due to
the increased matter density. These two body encounters will shift
the particles to superior orbits where
they'll find turning points, and bounce.\\

But even if an inner black hole forms, the influence upon the
accreting matter that it would exert would be limited by the
amount of mass it has compared with the mass of the non-baryonic
object (also note that the boson star radius and the shells where
most of the non-baryonic matter is located, are farther away than
the Schwarzschild radius of the presumed black hole by a factor of
at least 100). In general, there will be situations in which the
accretion rate will be so low, that even if a black hole is formed
with all the mass accreted during the lifetime of the universe, it
will still have several orders of magnitude less than the boson
star mass. Consider the center of the galaxy, its mass is believed
to be above 2 10$^6 M_\odot$, while the accretion rate is $\sim
10^{-6} M_\odot$ yr$^{-1}$. Then, in the absolutely worse case, if
a black hole is formed with the accreted mass in a period of
10$^{10}$ yr, it will have 10$^4 M_\odot$, and its gravitational
influence will be defied by the non-baryonic object. Boson stars
containing fermion objects within have been considered in the past
\cite{BOS-FERMION}, and this appear to be an extreme case where
the fermion (neutrons) component have collapsed to a
black hole. \\

Finally, a detailed analysis of the evolution of boson stars
subject to continuous inflow of non-baryonic particles was carried
out in Ref. \cite{Suen}. Their results showed that under finite
perturbations, the stars on the stable branch will settle down
into a new configuration with less mass and a larger radius. Then,
the accretion of non-baryonic matter possibly entering into the
condensate would not pose a problem to boson star stability, nor
generate a collapse. We recall that the instability of a boson
star with respect to gravitational collapse has been studied by
Kusmartsev et al. (see Ref. \cite{kus91}, see also Ref.
\cite{Heusler}), using catastrophe theory. Rotating boson stars
were studied (among others) by Schunck et al. \cite{rot}.\\

In summary, even when the physical model used here could be
regarded as too simplified and not complete (the star is not
rotating, and a mechanism by which diverting matter from the
center is not specifically given) we believe it still is important
to begin to address the issue of real astrophysical scenarios, as
accretion, upon theoretically foreseen non-baryonic objects. One
of the first steps in this direction is presented in this work.

\section{Steady and geometrically thin accretion disc}

A test particle rotating around an spherically symmetric object,
which generates a space-time described by the metric \be \label{m}
ds^2 = e^{\nu (r)} dt^2 - e^{\mu (r)} dr^2
  - r^2 ( d\vartheta^2 + \sin^2\vartheta \, d\varphi^2) \;,
\ee would do so at a velocity  $ v_\phi^2 = r \nu^\prime e^\nu/2$
\cite{Weimberg}.\footnote{This form of the velocity (and the
metric) may not be generally valid in the region dubbed as ``dark
matter zone" in the paper by Nucamendi et al. \cite{NUC}, i.e.
where the flat rotation curves of galaxies are found.} A prime
stands for the derivative with respect to the radial coordinate.
Let us call, for the ease of notation, $B=e^\nu$, then
$B^\prime/B=\nu^\prime$. In the case of a black hole, $B$ is given
by the Schwarzschild metric $B(r)=1-2M/r$, what finally yields the
Keplerian velocity $ v_\phi^2 = M/r $.  The angular velocity,
defined as usual as $\Omega=v_\phi/r$, is then proportional to
$r^{-3/2}$ and it decreases rapidly far from the center. For a
more generic case, the rotational velocity and its derivative will
be given by \be \Omega = \sqrt{ \frac{B^\prime}{2r} },
\hspace{2cm} \Omega^\prime = \left(
\frac{B^\prime}{2r}\right)^{-1/2} \left[
\frac{B^{\prime\prime}}{2r} - \frac{B^\prime}{2r^2} \right].\ee

We now turn to dimensionless quantities. The radial coordinate
being $x=mr$, where $m$ will be associated with the mass of the
boson. Then, $ v_\phi^2=dB/dx \; {x}/{2}.$ Recall that $c$ is
being implicitly taken equal to 1, the
dimensionfull velocity will just be $v_\phi \times c$.
Equivalently, the dimensionfull rotational velocity will be
$\Omega = m\; c\; \sqrt{ [(dB/dx) x/2 ]} $, and\be \Omega= \sqrt{
\frac{dB(x)}{dx} \frac x2} m[{\rm GeV}] \frac{1.5228 \times
10^{24}}{{\rm s}},\;\;\;\; \Omega^\prime = \frac {d\Omega}{dx}
m^2[{\rm GeV}] \frac {7.72973 \times 10^{42}}{{\rm km} \; {\rm s}
}.\ee

The power per unit area in the black body disc we are studying is
given by \cite{FKR} \be D(r) = - \frac { \dot M \Omega
\Omega^\prime }{4\pi} r \left[ 1 - \left( \frac{R_i}{r} \right)^2
\left( \frac{ \Omega_i }{\Omega} \right) \right], \label{D} \ee
where $R_i$ is the interior limit of the disc and $\Omega_i$ is
the velocity there. $\dot M$ is the accretion rate, it is an
external parameter that should be observationally obtained, $R_i$
should also be either observationally derived or theoretically
estimated, by taking into account the nature of the central
object, all other quantities are known if the explicit form for
the metric coefficients is. As we see from Eq. (\ref{D}), the
standard model for a steady, geometrically thin and optically
thick disc predicts a surface emissivity that is independent of
viscosity, given all other disc parameters. In fact, Eq. (\ref{D})
is strictly true only when the gravitational potential is $\sim
GM/r$, otherwise relativistic corrections would appear. The
benchmark case for disc accretion is that of a Schwarzschild black
hole. For it, $M$ is independent of $r$, it is just a constant
equal to the black hole mass, $M_{BH}$, and it is easy to
analytically derive the form for the rotational velocity and its
derivative by using the expressions given above. With these
functions, the power per unit area for a central black hole of
mass $M_{BH}$ is \be D(r)^{BH} = \frac{3}{8\pi} M_{BH,a} \left[ 1-
\left(\frac{x_i}{x} \right)^{1/2} \right] x^{-3} \dot M
[M_\odot/yr] \; m^2[{\rm GeV}] \; 1.47065 \times 10^{74} \frac
{{\rm erg}}{{\rm cm}^2 {\rm s}}.\ee The appearance of $m$ in the
latter formula is justified because of the use of the
dimensionless radial coordinate and the definition of $M_{BH,a}$
by means of the equality $ M_{BH}=M_{BH,a} \;\; {M_{pl}^2}/{m},$
where $M_{pl}$ is the Planck mass. This particular form of writing
the hole mass is useful because it will allow a direct comparison
with the case of a boson star. In a general situation (when not
necessarily a black hole is the central object), the power per
unit area will be given by \be D(r)^{BS} = -\frac{1}{4\pi}
[\Omega] [\Omega^\prime] \left[ 1- \left(\frac{x_i}{x} \right)^2
\left(\frac {\Omega_i}{ \Omega} \right) \right] x \dot M
[M_\odot/yr] \; m^2[{\rm GeV}] \; 1.47065 \times 10^{74} \frac
{{\rm erg}}{{\rm cm}^2 {\rm s}}.\ee Here, $[\Omega]$ and
$[\Omega^\prime]$ stand for the dimensionless form of these
quantities. The superscript $BS$ refers to our later use
of this formula for the case of a boson star.\\

If the disc is optically thick, as we are assuming, the local
effective temperature must be sufficient to radiate away the local
energy production. The temperature of the disc is then related
with the latter result by $ D(r)= \sigma T^4$, as it is
appropriate for a black body system; $\sigma$ being the
Stefan-Boltzmann constant (equal to $5.67 \times 10^{-5} $ erg
s$^{-1}$ cm$^{-2}$ K$^{-4}$). For the luminosity, $L(\nu)$, and
flux, $F(\nu)$, we shall use the relationship \be L(\nu)= 4 \pi
d^2 F(\nu) = \frac {16 \pi^2 h \cos i \nu^3}{c^2}
\int_{R_i}^{R_{out}} \frac{r dr}{ e^{h\nu/kT} - 1 } \; , \ee where
$d$ is the distance, $R_{out}$ is the outer border of the disc,
$i$ the disc inclination, and  $h (=6.6256 \times 10^{-27}$ erg s)
and $k (=1.3805 \times 10^{-16}$ erg K$^{-1}$) are the Planck and
Boltzmann constants, respectively. In order to make the first
comparison we shall take $ R_i = 3 R_g = 3 \times 2GM = 6
M_{pl}^{-2} M_{pl}^{2} \frac{M_a}{m}\Rightarrow x_i=6 M_a,$ and $
R_{out} = 30-50 R_g = 30-50 \times 2GM = 60-100 M_{pl}^{-2}
M_{pl}^{2} \frac{M_a}{m} \Rightarrow x_{out}=60-100 M_a, $ with
$M_a$ being the dimensionless mass. This will let us know whether,
even in the non-relativistic regime and for the same initial
commencement of the disc, we are able to see any difference in the
emitted spectrum.
The relativistic analysis will follow in the next section.\\

In dimensionfull form, for an inclination of 60 degrees, changes
of variables yield a luminosity equal to, \be L(\nu)= 2.2559
\times 10^{-73} {\rm erg}\; m^{-2}[{\rm GeV}]\; \nu^3[{\rm
s}^{-3}]\; \int_{x_i}^{x_{out}} \frac{x dx}{ e^{h\nu/kT(x/m)} - 1
} \; . \ee $e^{h\nu/kT(x/m)}$ must be obtained for each particular
central object, through the solution for the metric coefficients
and the computation of the rotational velocities and temperatures
(for the latter we use $D(r)=\sigma T^4$, with the corresponding
power) We shall now apply the presented formulae to the case of a
boson star. For readers not familiar with the way in which boson
star configurations are obtained, we suggest Refs.
\cite{MIELKE,reviews,col86,TORRES-BOS,kus91,RB,KAUP,rotating}.

\section{Numerical results for a non-relativistic disc}

We can now note the following basic steps of the computation of
the emission properties of the disc: 1) Define the boson mass $m$,
the inner and outer borders of the disc and the accretion rate
$\dot M$; 2) Compute $e^{h\nu/kT(x/m)}$ for different frequencies;
3) Integrate $\int_{x_i}^{x_{out}} \frac{x dx}{ e^{h\nu/kT(x/m)} -
1 }$ and obtain the luminosity for a given frequency; 4) Repeat
the previous steps for several frequencies and build up the
spectrum. We shall take a model with central mass equal to $2.5
\times 10^6 M_\odot$. \footnote{Since we have already assumed that
$\Lambda=0$, for a maximal boson star with central density
$\sigma(0)=0.271$, we shall need the boson
mass to be equal to 
$\sim 10^{-26}$ GeV. A discussion of this value of $m$ on the
light of particle physics was given in Ref. \cite{GALAXY}, see
also Ref. \cite{MIELKE}. In any case, note that because of the
scaling properties, the general trend of the results presented
here is valid  for every value of $m$, even for those leading to
stellar sized boson stars. This $\sim 10^{-26}$ GeV value is
obtained equating the boson star mass, $M_{BS}=0.633\;M_{pl}^2/m$
to the mass of the central object, where, in the previous
equation, $0.633 \equiv M_{BS,a}$ is the value of the boson star
mass in dimensionless units; this value is numerically obtained
solving the star equations of structure. In Ref. \cite{GALAXY}, we
presented a relation between the boson mass $m$ expressed in GeV
and the total mass of the star expressed in millions of solar
masses. This relationship is $ m[{\rm GeV}]\simeq 1.33 \times
10^{-25} {M_{BS,a}}/{M[10^6 M_\odot]}, $ and was used to obtain
the value of  $m$ quoted above.} We shall take $\dot M$ to be
equal to a $few\; \times 10^{-6} M_\odot$yr$ ^{-1}$; in the
computation that follows, if not otherwise stated, the pre-factor
will be taken as 2. In this situation, $x_i$ will be equal in this
case to
3.798 and $x_{out}=(30-50)\; x_g$, with $x_g=1.266$. \\

{\it Remark on scaling:} Note that if we would be interested in
other different masses for the compact object, we would just need
to change the boson mass. All quantities are already scaled, so
that a change in $m$, and eventually in $\dot M$, will readily
give the values of all other
physical results.\\

\begin{figure}[t]
\vspace{-1.5cm}
\begin{center}
\includegraphics[width=8cm,height=11cm]{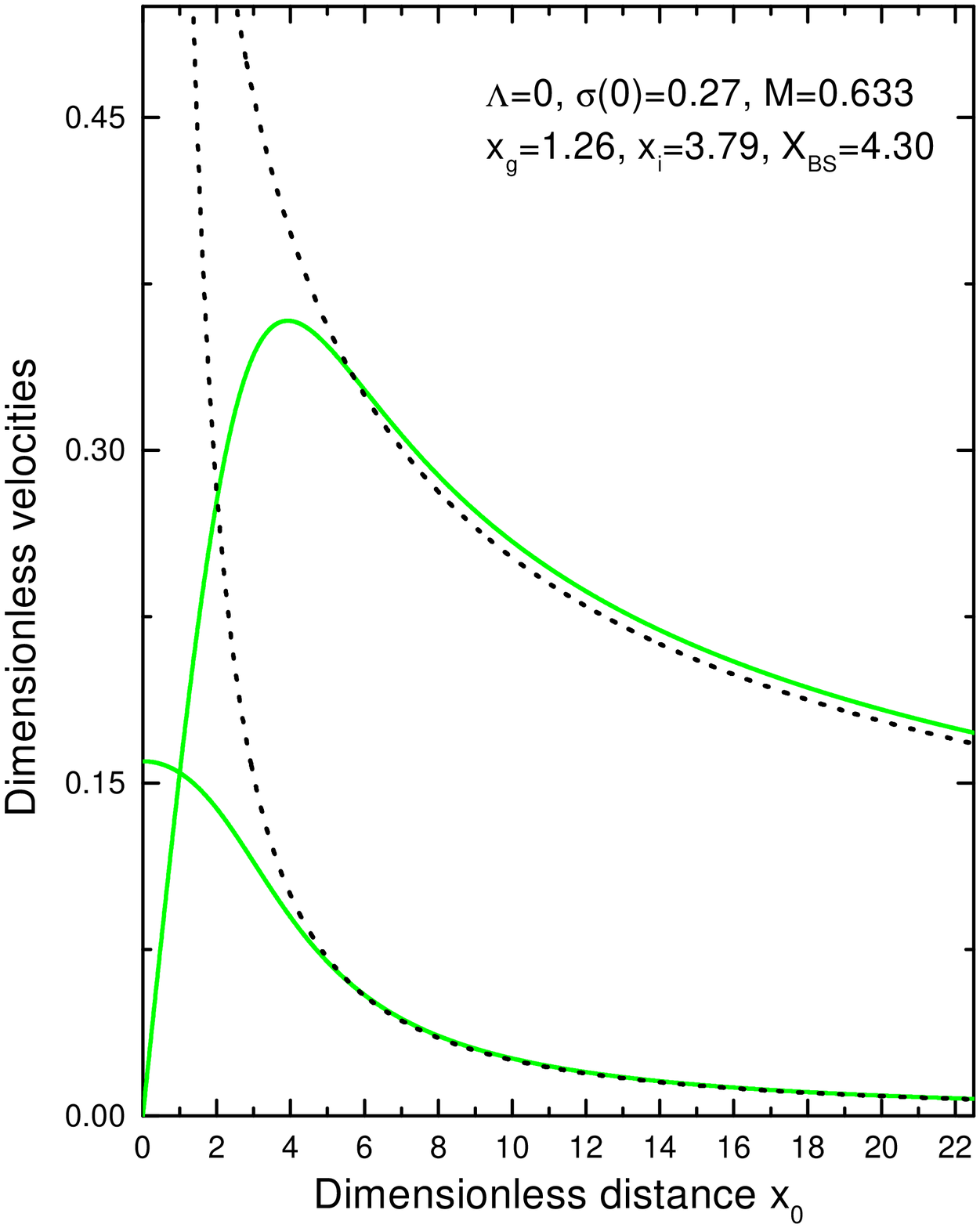}
\includegraphics[width=8cm,height=11cm]{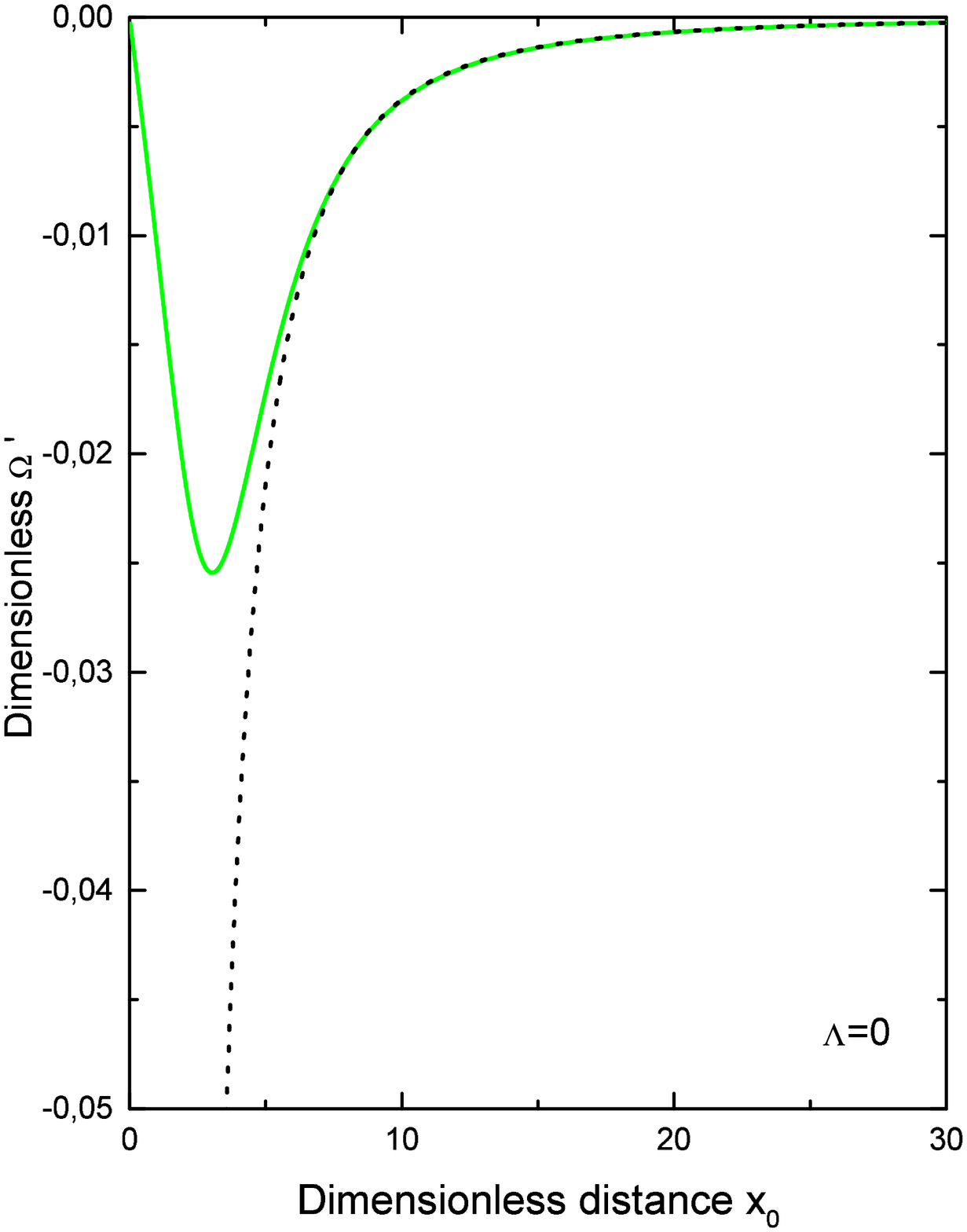}
\end{center}
\vspace{-1.5cm} \caption{Left: Dimensionless velocities. The upper
set of curves are $v_\phi$, the lower one are the rotational
velocity $\Omega$, both for a boson star (solid) and a black hole
(dash). The boson star has a dimensionless radius equal to
$x_{BS}=4.3$, and a dimensionless mass equal to 0.633. The black
hole has equal mass. Right: Dimensionless derivative of the
rotational velocity, $\Omega^\prime$.} \label{l01}
\end{figure}

\begin{figure}[t]
\vspace{-1.5cm}
\begin{center}
\includegraphics[width=8cm,height=11cm]{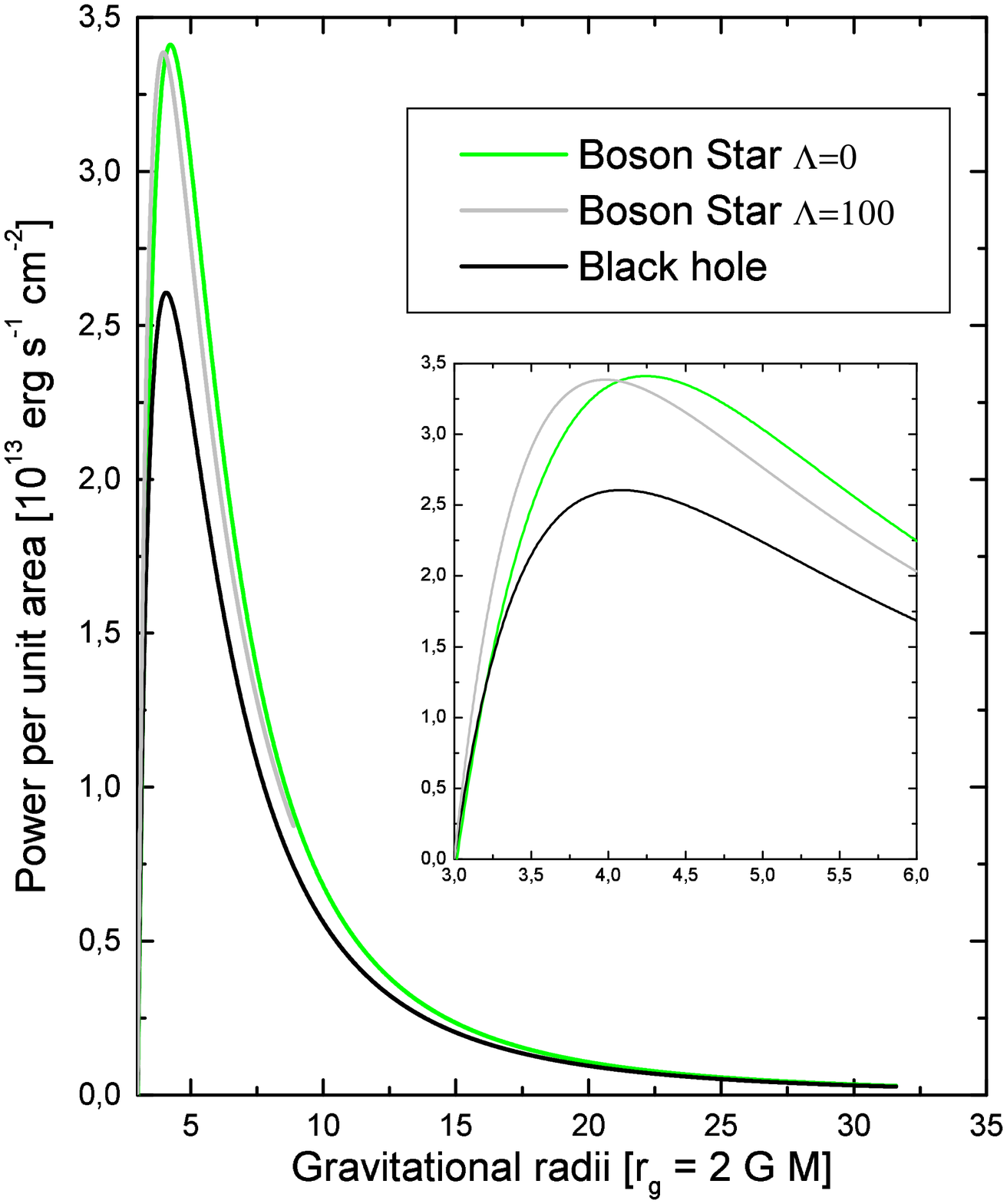}
\includegraphics[width=8cm,height=11cm]{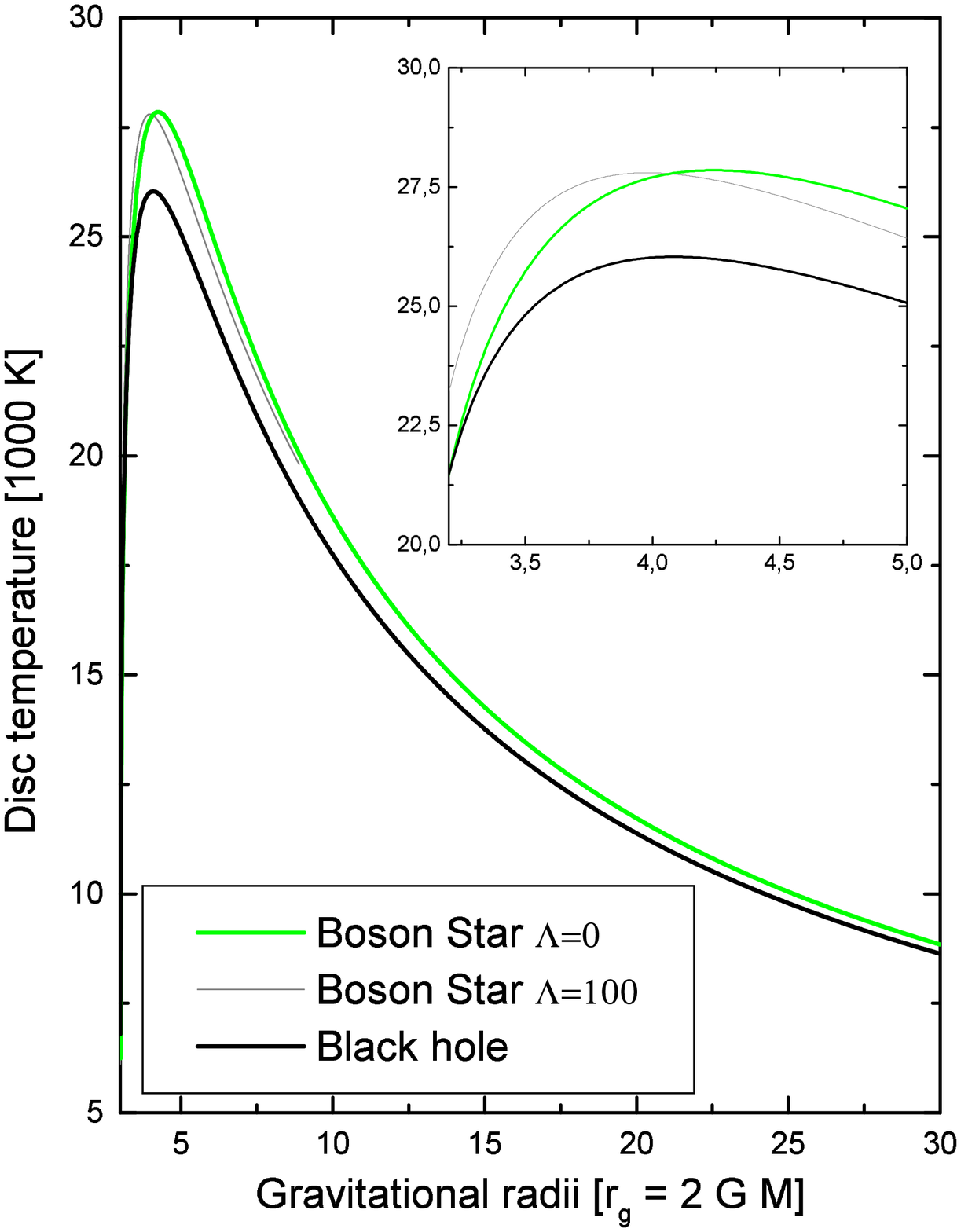}
\end{center}
\vspace{-1.5cm} \caption{Left: Dimensionfull power per unit area
for a black hole, and boson stars with self-interaction equal to 0
and to 100, all of the same mass. The position where the maximum
of the curves are is zoomed out in the figure. Right:
Dimensionfull temperatures of the disc for a black hole, and boson
stars with self-interaction equal to 0 and to 100, all of the same
mass. The position where the maximum of the curves are is zoomed
out in the figure. } \label{DPT}
\end{figure}

In Figure \ref{l01} we show the dimensionless rotational
velocities, both $v_\phi$ and $\Omega$, for the previously quoted
model. Through the comparison of the rotational velocities that a
black hole of equal mass would produce, it is easy to see that a
boson star is a highly relativistic object. The velocities for a
black hole center are also shown in the Figure. A comparison of
the power per unit area is shown in Figure \ref{DPT}. It is also
shown there the difference between the temperatures of the disc.
As one would expect from the behavior of the metric coefficients,
the boson star curves, especially farther from the central object,
tend to mimic those of the black hole case. However, in the inner
parts of the disc, it is apparent that a slight deviation can be
noticed. The boson star produce more power per unit area, and a
hotter disc, than a black hole of equal mass. In order to see
whether the dependence on the self-interaction parameter is
strong, we have computed these same quantities for the case of
$\Lambda=100$ and the same total mass and accretion rate. We have
used a central density equal to $\sigma(0)=0.093$, what yields a
boson star mass -in dimensionless units- equal to 2.25. In order
for this dimensionless mass to rightfully represents the size of
the central object, we need a boson mass equal to $1.20 \times
10^{-25}$GeV.
\begin{figure}[t]
\vspace{-1.5cm}
\begin{center}
\includegraphics[width=8cm,height=11cm]{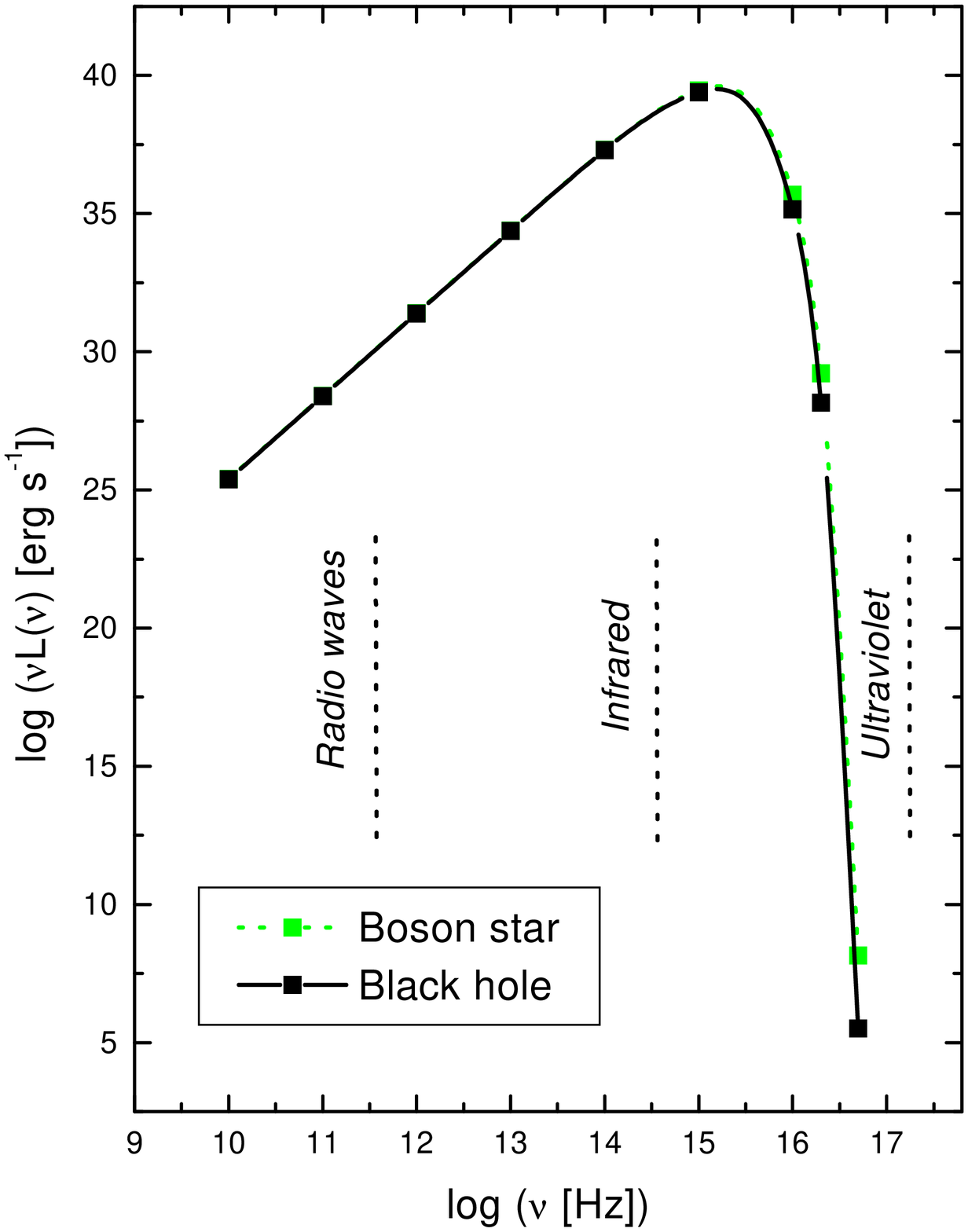}%
\end{center}
\vspace{-1.5cm} \caption{The final emission spectrum. Differences
between the boson star cases with and without self-interaction are
negligible, they are not noticed in the plot. Noticeable
deviations between a supermassive black hole and a boson star
begin at the far ultraviolet. } \label{spec}
\end{figure}
The final spectrum of the radiation emitted by the disc was
computed and it is shown in Figure \ref{spec}. It is clearly seen
that throughout most parts of the electromagnetic spectrum, the
obtained differences in the disc properties, such as the
temperature or the power per unit area, will not be
observationally noticed. Differences between the boson star cases
with and without self-interaction are negligible. However, it
appears that differences between black hole and boson stars
centers turn out to be important at the most energetic part of the
electromagnetic range, beginning in the far ultraviolet. The disc
model we are considering does not produce energy in the X-ray and
gamma-ray bands, but the possibility is open that even in the
simplest non-relativistic case we have considered, a boson star
and a black hole case could in principle be distinguished by the
radiation properties at high energies. We can expect that better
and more realistic models of accretion discs can show
stronger differences.\\

The metric potentials of a boson star make the rotational
velocities and their derivatives, except for the innermost parts
of the disc, always greater than their counterparts produced by a
central black hole. Looking at the left panel of Figure \ref{l01},
it can be seen that only for the more central coordinates, the
black hole rotational velocity is greater than that produced by a
boson star. From the right panel we see, in addition, that the
derivative of $\Omega$ is also always greater for a boson star
than for a black hole. Recalling that $D(r) \propto \Omega
\Omega^\prime$, a greater rotational velocity and rotational
velocity gradient implies a greater power. The same happens for
the temperature of the disc. The innermost part of Figure
\ref{l01}, where the rotational velocity of the particles rotating
around a black hole diverges, can not be taken into account. These
particles are all following unstable orbits and are not part of
the disc, which begins at $x_i=3x_g$, at a value of about 4 in the
x-axis of Figure \ref{l01}. The effect of a larger temperature and
power per unit area is then the integrated output of a disc
rotating faster when a boson star is the central object.
\footnote{We have been using central object and accretion disc
parameters, like mass and accretion rate, consistent with those
inferred for our own Galaxy \cite{Genzel}. We can then ask if the
replacement of the presumed central black hole for a boson star
can do any better in fitting the observationally obtained
spectrum. The answer to this question is no. However, this is not
(at this time) because of the fact of the different central
object, but because of the properties of the accretion disc
itself. The disc model here considered simply does not work for
Sgr A$^*$, neither with a black hole, nor with a boson star. The
measured $M$ and $\dot M$ for Sgr A$^*$ would yield -within the
standard theory of a steady, optically thick, geometrically thin
accretion disc- an accretion luminosity equal to 0.1$\dot
Mc^2>10^{40}$ erg s$^{-1}$, assuming a nominal efficiency of 10\%.
However the total luminosity of Sgr A$^*$ is less than $10^{37}$
erg s$^{-1}$. In addition, the value of $M$ would make the
standard accretion disc broad band spectrum to have its peak in
the infrared region, what is opposite to observation. The spectrum
is --with the exception of a few bumps-- essentially flat, with
detected flux even in x and $\gamma$-rays. A recent compilation of
luminosity measurements for Sgr A$^*$ was given by Narayan et al.
\cite{NARA}. Observations suggest, then, that Sgr A$^*$ is not
behaving like a blackbody, disregarding the nature of the central
object. An alternative solution for the blackness problem of Sgr
A$^*$ is the so-called advection dominated accretion discs, or
ADAF \cite{NARA,narayan95}. The key for ADAF models is that most
of the power generated by disc viscosity is advected into the
hole, while only a small fraction of it is radiated away. An
essential ingredient for ADAF models is then the existence of an
event horizon. We remark that for the less active galaxies, like
ours, a boson star seems to be a more problematic center than a
black hole: the blackness problem could
even be more severe. }\\

Before ending this section we shall provide a brief discussion on
the boson mass chosen. This discussion is based in our previous
paper \cite{GALAXY}, as well as in \cite{NEW}. As discussed in
\cite{GALAXY}, based on the constraints imposed by the mass-radius
relationship valid for the scalar stars analyzed, we may conclude
that 1. if the boson mass is comparable to the expected Higgs mass
(hundreds of GeV), then the center of the galaxy could be a
non-topological soliton star \cite{lee87}, 2. an intermediate mass
boson could produce a super-heavy object in the form of a boson
star, and 3. for mini-boson stars to be used as central objects
for galaxies the existence of an ultralight boson is needed. These
conclusions are to considered as order of magnitude estimates. If
boson stars really exist, they could be the remnants of
first-order gravitational phase transitions and their mass should
be ruled by the epoch when bosons decoupled from the cosmological
background. The Higgs particle could be a natural candidate as
constituent of a boson condensation if the phase transition
occurred in early epochs. A boson condensation should be
considered as a sort of topological defect relic. If soft
phase-transitions took place during cosmological evolution e.g.,
soft inflationary events, the leading particles could have been
intermediate mass bosons and so our super-massive objects should
be genuine boson stars. If the phase transitions are very recent,
the ultralight bosons could belong to the Goldstone sector giving
rise to miniboson stars. We should also mention the possible
dilatons appearing in low-energy unified theories, where the
tensor field of gravity is accompanied by one or several scalar
fields. In string effective supergravity, the mass of the dilaton
can be related to the supersymmetry breaking scale $m_{SUSY}$ by
$m_\phi \sim 10^{-3} (m_{SUSY} / TeV)^2$ eV. Finally, a scalar
with a long history as a dark matter candidate is the axion, which
has an expected light mass $m_\sigma \sim 7.4 (10^7$ GeV$/
f_\sigma )$eV $\sim 10^{-11}$ eV with decay constant $f_\sigma$
close to the Planck mass.

\section{Coordinate dependent Eddington luminosity}

\begin{figure}[t]
\vspace{-1.5cm}
\begin{center}
\includegraphics[width=8cm,height=11cm]{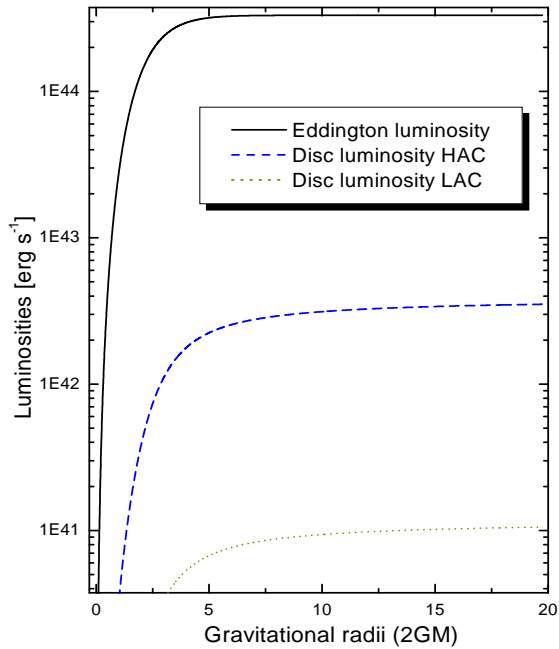}
\end{center}
\vspace{-1.5cm} \caption{Comparison between the dimensionfull
Eddington luminosity, now dependent on the radial coordinate, and
the power radiated by the disc at different radius. HAC stands for
a high accretion rate, here taken to be equal to  $2\times 10^{-4}
M_\odot$yr$^{-1}$. LAC stands for a low accretion rate, equal to
$2\times 10^{-6} M_\odot$yr$^{-1}$. The coordinate dependent
Eddington luminosity is everywhere greater than the power
generated by a thin disc about a boson star. Note that the form of
the curve for the Eddington luminosity is provided by that of the
mass distribution within the boson star.} \label{edd}
\end{figure}

It is interesting to make now a comment on the Eddington
luminosity ($L_E$) when a boson star is the central object.
Consider first the basic concept. If the luminosity produced by
the accreted material is too great, then the radiation pressure it
would produce would blow up the infalling matter. The limiting
luminosity, known as $L_E$, is found by balancing the inward force
of gravity with the outward pressure of radiation \cite{FKR}. This
limit is then found by assuming that the infalling matter is fully
ionized, and that the radiation pressure is provided by Thompson
scattering of photons against electrons -which in turn are
strongly coupled with the protons resident in the plasma by the
electromagnetic interaction. Equating the involved forces results,
in a first approximation, $ {\sigma_T L}/{4 \pi r^2
c}={GMm_p}/{r^2},$ where $\sigma_T$ is the Thompson scattering
cross section and $m_p$ is the proton mass. When this relationship
is fulfilled, $L= L_E$. $L_E$ is the maximum luminosity which a
spherically symmetric source of mass $M$ can emit in a steady
state. Longair \cite{LONGAIR} provides the value of the Eddington
luminosity in watts by introducing the gravitational radius $r_g$,
$L_E= {2 \pi r_g m_p c^3}/{\sigma_T}=1.3 \times 10^{31}
{M}/{M_\odot} {\rm W}. $ None of the active galactic nuclei were
found to exceed this limit \cite{LONGAIR}. It is interesting to
note that if $M$ is constant, as in the case of a black hole, this
limit is independent of the radius, that is, if $L<L_E$ at any
given radius, the inequality will be sustained everywhere. A boson
star -as well as any other transparent object- has a non-constant
mass distribution and the Eddington luminosity become a coordinate
dependent magnitude.\footnote{A boson star could well turn into a
non-transparent object if we are to admit the possibility of
scalar electrodynamics effects, where considerable cross sections
with photons may appear. This is an interesting problem, yet to be
attacked, of boson star physics.} In this situation, one may ask
if there is any point within the stellar structure such that, even
if $L<L_E$ outside the star, the opposite inequality is valid at
this point. If this is so, and $L>L_E(r=r_0)$, steady accretion
would not proceed. We would so be able to define the internal
border of the disc. $L_E(r)$ is given by the distribution of mass,
since $L_E(r) \propto M(r)$. For the luminosity produced by the
disc we have to integrate the power produced by a disc slab of
size $dr$ from $r_i$ to $r$. This is what should be compared with
$L_E(r)$. Then, we compute the total power radiated between $r_i$
and $r$ as \be P = 2 \int_{r_i}^r D(r) 2\pi r dr= -5.705 \times
10^{46} \; \dot M [\frac{M_\odot}{ {\rm yr}}]\; \frac{{\rm
erg}}{{\rm s}^{-1}}\; \int_{r_i}^r \frac 12\;
[\Omega]\;[\Omega^\prime]\; x^2 dx.\ee  In Figure \ref{edd}, we
show the results for the Eddington luminosity for a non
self-interacting boson star, compared with the power radiated by
the disc for two different accretion rates. In none of these
cases, the power generated in circles whose radii are smaller than
the star size overcomes $L_E$. This also supports the idea that
accretion can continue inwards, within the boson star structure.

\section{A relativistic treatment for the accretion}

Even when useful as a first approach, the presented treatment fail
when considering the innermost part of the disc, especially when
the disc orbiting a boson star has a more internal commencement
than that orbiting around a black hole. In order to get a complete
picture, we have to treat the problem in a relativistic way, this
is objective of the rest of this paper.

\subsection{Particle orbits}

Consider again the stationary, spherically symmetric, time
independent, line element, written as \be ds^2= -B(r) dt^2 +
\frac{1}{1-2M(r)/r}dr^2 + r^2 ( d\theta^2 + \sin^2\theta \,
d\phi^2) \;. \label{1}\ee Again, the Schwarzschild solution is a
special case of Eq. (\ref{1}), occurring when $B(r) = 1-2M(r)/r$
and $M(r)$ is a constant equal to the mass of the black hole. In
our present situation, $B(r)$ must be consistent both with an
asymptotically flat space-time and with the absence of a
singularity. $M(r)$, in turn, decreases with decreasing radius,
being equal to 0 at $r=0$. The particle orbits in that general
metric will be determined by the conserved quantities, $ E=-p_t:
{\rm total\; energy} $; $L=p_\theta: {\rm component\;of\; the\;
angular\; momentum.}$ An additional constant of motion will be the
mass of the test particle, let us call it $\hat m$, which can be
absorbed by redefining quantities in a per-unit-mass basis. The
general (non-trivial) equations for the orbital trajectory in a
space-time described by Eq. (\ref{1}) are: \be r^2 \left[g_{rr}
g_{tt}\right]^{1/2} \frac {dr}{d\lambda} = V(r)^{1/2}= \left[ E^2
r^4 - r^4 B \left(1+ \frac{L^2}{r^2}\right) \right]^{1/2}, \ee \be
r^2 \frac{d\phi}{d\lambda} = \frac{L}{\sin^2\theta}. \ee For a
general derivation see the book by Weinberg \cite{Weimberg} (take
caution with the different definition of the symbols). Here,
$\lambda$ is an affine parameter related with the proper time
$\tau$, by $\lambda=\tau/ \hat m$. We shall take $\hat m =1$ for
massive particles. A prime
will denote derivation with respect to $r$.\\
\begin{figure}[t]
\vspace{-1.5cm}
\begin{center}
\includegraphics[width=8.5cm,height=10cm]{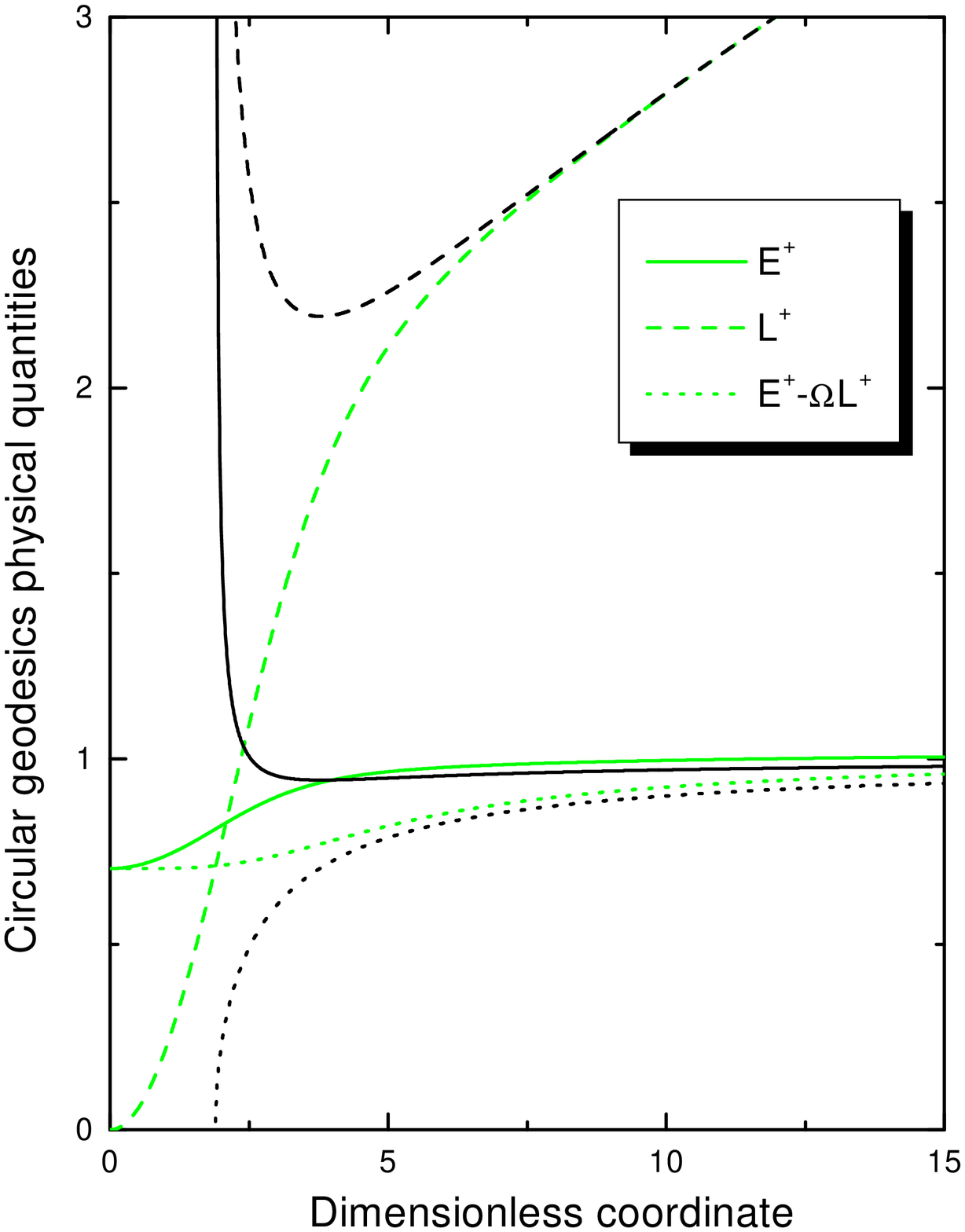}
\includegraphics[width=8.5cm,height=10cm]{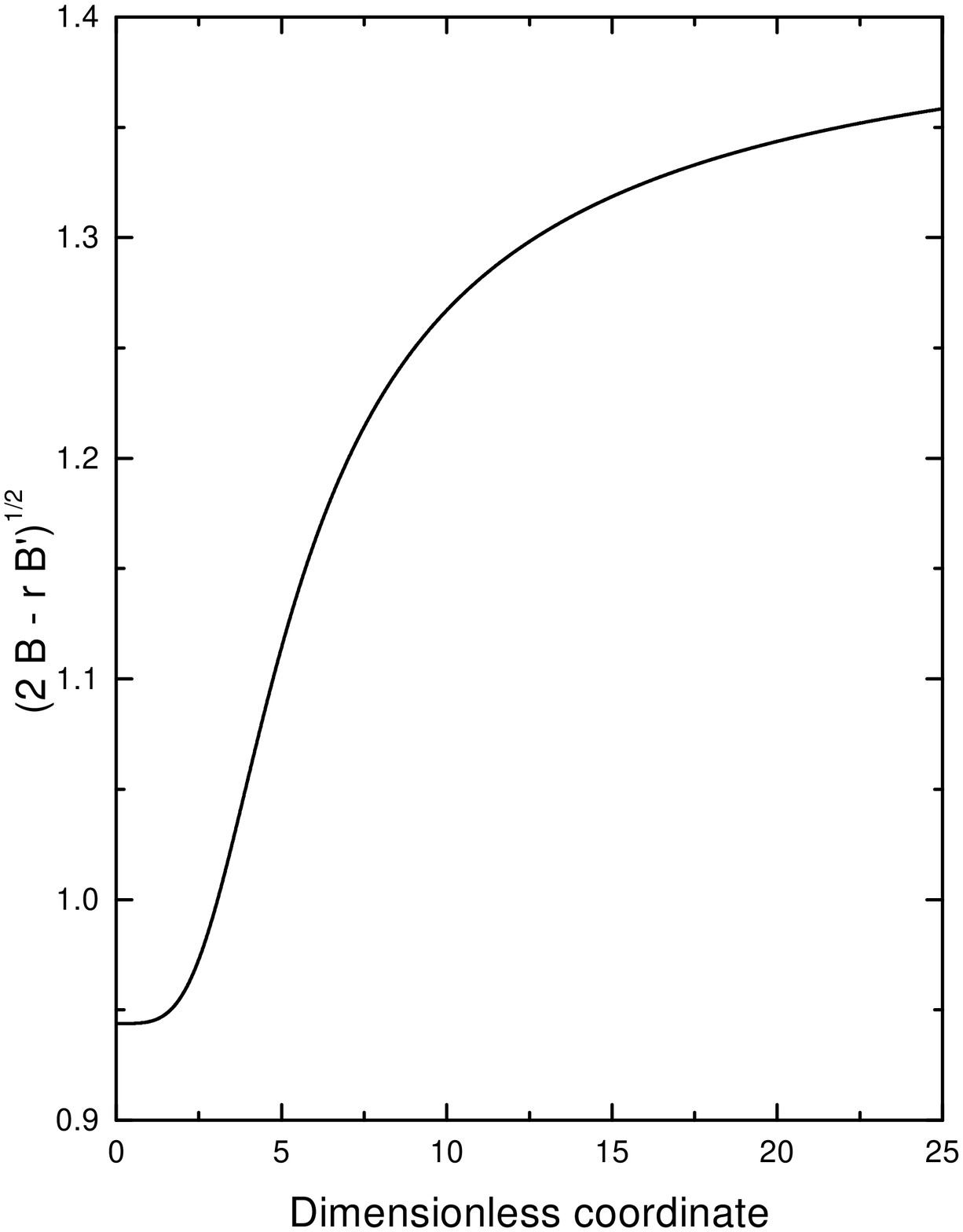}
\end{center}
\vspace{-1.5cm} \caption{Left: Radial dependence of the energy,
angular momentum, and the difference between them -with the
momentum weighted by the rotational velocity $\Omega$- for
circular geodesics, all per unit mass. The divergent curves are
the corresponding magnitudes for the Schwarzschild black hole
case. Right: Radial dependence of the denominator of the
expressions for $E^\dag$ and $L^\dag$. Being the values it takes
all positive, it is shown that it is possible to find a circular
orbit even within the boson star. Details of the model are as
follows: The mass for both, the black hole and the boson star, was
taken as $M=0.633 m_{\rm Pl}^2 / m$, with $m$ the mass of the
constituent bosons, a free parameter. The central density, in
usual dimensionless units, is equal to 0.27. No self-interaction
was considered. All figures in this Section are based on this same
model, but the behavior is generic. The radial coordinate is
$x=mr$. } \label{circ}
\end{figure}

We are interested in the circular orbits. For such class,
$dr/d\lambda$ must vanish instantaneously and at all subsequent
moments, what imposes the conditions, $ V(r)=0\;\;\; {\rm
and}\;\;\; V(r)^\prime=0 .$ These equations determine $E$ and $L$,
which per unit mass, are given by \be \frac{E}{\hat m}=E^\dag=
\left( \frac{2B^2}{2B-rB^\prime}
\right)^{1/2},\;\;\;\;\;\;\frac{L}{\hat m}=L^\dag= \left(
\frac{B^\prime r^3}{2B-rB^\prime} \right)^{1/2}.  \label{daga}\ee
These quantities are shown --both for the black hole and the boson
star case-- in the left panel of Figure 5. One can immediately
show that both $E^\dag$ and $L^\dag$ reduce themselves to the
Schwarzschild expressions when $B(r)=1-2M/r$, and that they tend
to them asymptotically. Circular orbits might not exist for all
values of $r$. It is needed that the denominator of $E^\dag$ and
$L^\dag$, be well-defined, i.e. $2B-rB^\prime
> 0$. But in the case of a boson star, this happens for all values
of the radial coordinate, see the right panel of Figure 5. Then,
{\it in a relativistic non-baryonic potential, generated by a
non-rotating compact object, there are circular orbits for every
possible value of the radial coordinate, including those which are
inside the structure.}  It is interesting to note, in addition,
that $E^\dag < 1$, so there
are no unbound orbits, hyperbolic in energetics \cite{BARDEEN}. \\

The circular orbits might not all be stable. Stability requires
that, when evaluated using the general expressions given in Eq
(\ref{daga}), $V(r)^{\prime\prime}\leq 0$. This yields \be
V(r)^{\prime\prime} (2 B -r B^\prime) = -6 B B^\prime r^3 + 4
(B^\prime)^2 r^4 - 2B B^{\prime\prime} r^4 \leq 0 .\ee Again, this
reduces to the Schwarzschild case when $B(r)=1-2M/r$. The results
of this computation in a generic boson star potential are shown in
the left panel of Figure 6. For all values of $r$,
$V(r)^{\prime\prime}\leq 0$. Then, {\it in a relativistic
non-baryonic potential, generated by a non-rotating compact
object, all circular orbits, even those within the structure, are
stable.}\footnote{An alternative derivation of the previous
results can be obtained as follows. Consider again the equation of
motion for $r$, written in the following slightly modified form:
\be \dot r ^2 = \frac{1}{g_{rr}g_{tt}} E^2 - \frac {1}{g_{rr}}
\left(1+ \frac{L^2}{r^2} \right).\ee  By multiplying both sides of
the latter equation by $g_{rr}g_{tt}$, and defining a new radial
coordinate by $ z=\int \sqrt {g_{rr}g_{tt}} dr ,$ we get $ \dot
z^2 = E^2 - 2V_{eff}(r).\label{3}$ We have considered that
$r=r(z)$, being the effective potential defined by \be
V_{eff}=\frac 12 B(r) \left(1+ \frac{L^2}{r^2} \right). \ee This
potential is depicted, for different values of $L/M$ (with $M$
being the total mass of the object in dimensionless units) in the
right panel of Figure 6. Classical mechanics allows to extract, in
the usual way, the orbital behavior. Superposed in the same plot
of Figure 6, we have also depicted the black hole effective
potential.
Differences between them arise just from the $rr$-metric
coefficient $B(r)$, and are noticed in the innermost regions. A
particle with any given energy can, in the non-rotating boson star
vicinity, be in an stable, circular orbit, at any value of the
radial coordinate. Depending on its energy, it can encounter one
or two turning points, but there is no capture trajectory,
consistently with the absence of singularity.}\\
\begin{figure}[t]
\vspace{-1.5cm}
\begin{center}
\includegraphics[width=8.5cm,height=10cm]{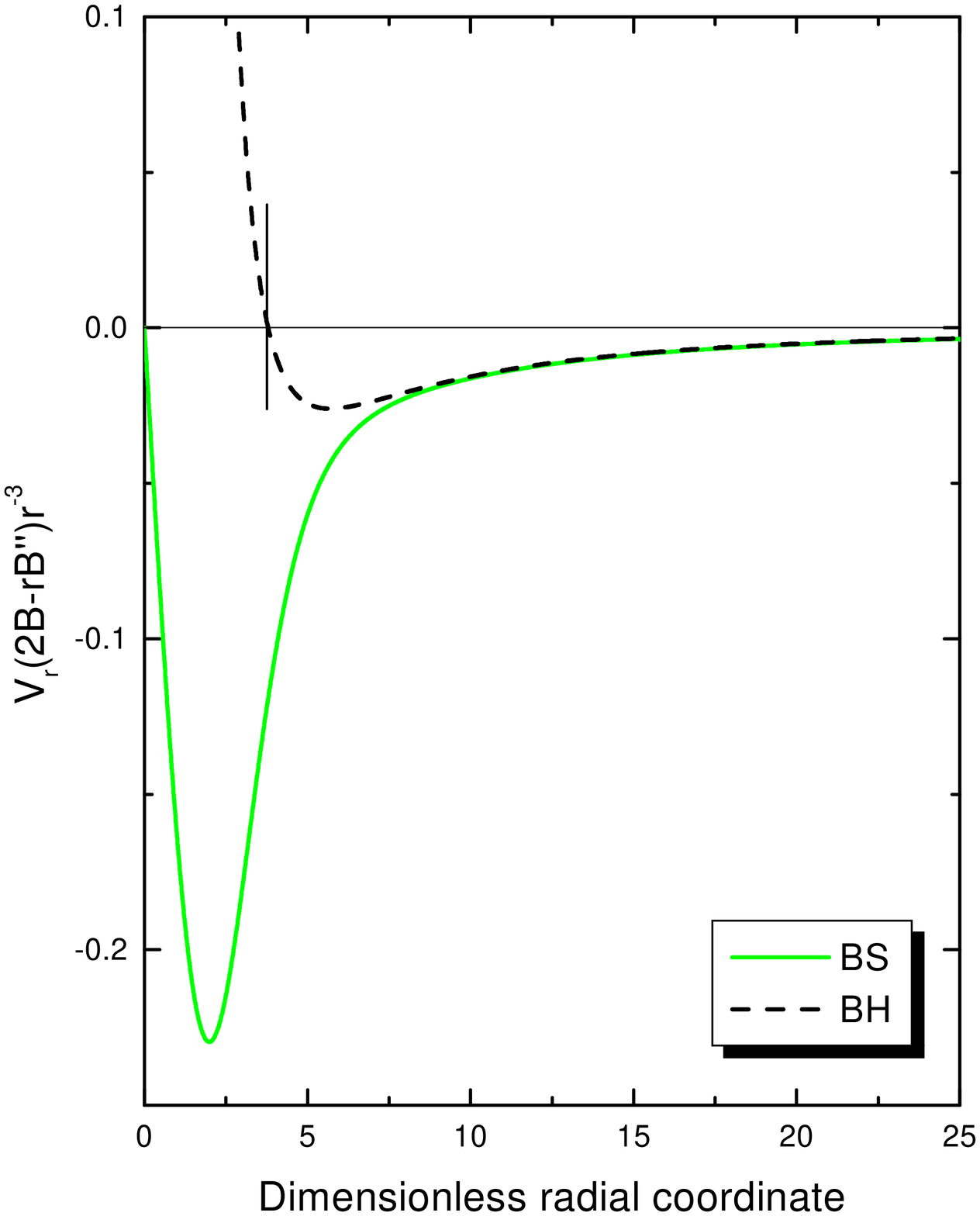}
\includegraphics[width=8.5cm,height=10cm]{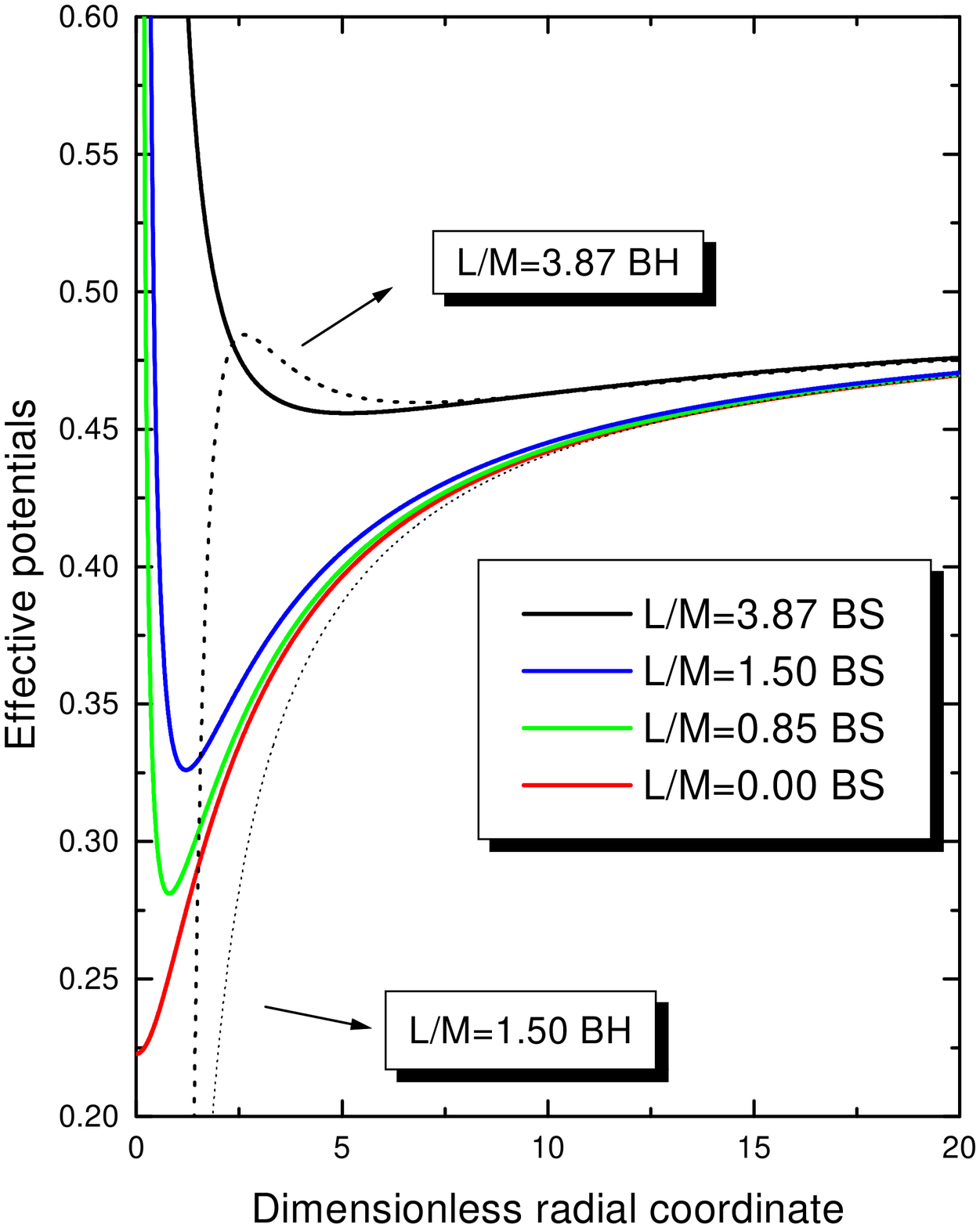}
\end{center}
\vspace{-1.5cm} \caption{Left: Radial dependence of the second
derivative of $V(r)$, both for a Schwarzschild black hole (dash
line) and a boson star (solid line).  The minimum radius at which
a stable orbit exists in the black hole case is $r=6M$, it is
marked with a crossing in the plot. For the boson star, all
circular orbits are stable. Right: Effective potential for
different values of $L/M$, with $M$ the dimensionless total mass
of the star, as in Figure 5. Two cases of the black hole effective
potential are shown for comparison. BS(H) stands for boson star
(black hole). The $x$-axis is the $z$ variable, introduced in the
text. For the case of a black hole, $z$ and $x$ coincide. }
\end{figure}

\subsection{From relativistic orbits to relativistic spectrum}

The power per unit area generated in a disc rotating around a
compact central object, which produces a relativistic potential,
is given by \cite{PAGE} \ben  D(r)&=&\frac {\dot M}{4\pi} \frac 1r
\left[ B\left( 1-\frac{2M}{r} \right) \right]^{1/2} \left
(-\frac{d\Omega}{dr} \right) \left( E^\dag - \Omega L^\dag
\right)^{-2} \int_{r_{ms}}^r \left( E^\dag - \Omega
L^\dag \right) \left (\frac{dL^\dag}{dr} \right) dr = \nonumber \\
&& 1.4706 \times 10^{74} \frac{{\rm erg}}{{\rm cm}^{2} {\rm s}}
m^2[{\rm GeV}] \frac {\dot
M[M_\odot {\rm yr}^{-1}]}{4\pi} \times \hspace{8cm} \nonumber \\
&&\frac 1x \left[ B(x)\left( 1-\frac{2M(x)}{x} \right)
\right]^{1/2} \left (-\frac{d\Omega}{dx} \right) \left( E^\dag -
\Omega L^\dag \right)^{-2} \int_{x_{ms}}^x \left( E^\dag - \Omega
L^\dag \right) \left (\frac{dL^\dag}{dr} \right) dx. \een  All
physical quantities are to be used in dimensionless form (e.g. $B
\longrightarrow B(x)$), and are explicitly given by \be E^\dag=
\left ( \frac {2B^2}{ 2B-xB^\prime} \right)^{1/2} , \hspace{3cm}
\Omega L^\dag= \left ( \frac {x B^\prime}{ 2 (2B-xB^\prime ) }
\right)^{1/2} .\ee

Clearly, when comparing with the output produced by a disc
rotating upon a central black hole, $D(r)$ will change not only
because of the modifications in $\Omega$, $E^\dag$, and $L^\dag$,
but also because of the change in the integration limits. Here,
$r_{ms}$ is the position of the innermost stable circular geodesic
orbit. Using the latter expression and all obtained results for
other physical parameters of the particle orbits, we can get the
relativistic results we were searching.\\

\begin{figure}[t]
\vspace{-1.5cm}
\begin{center}
\includegraphics[width=8.5cm,height=10cm]{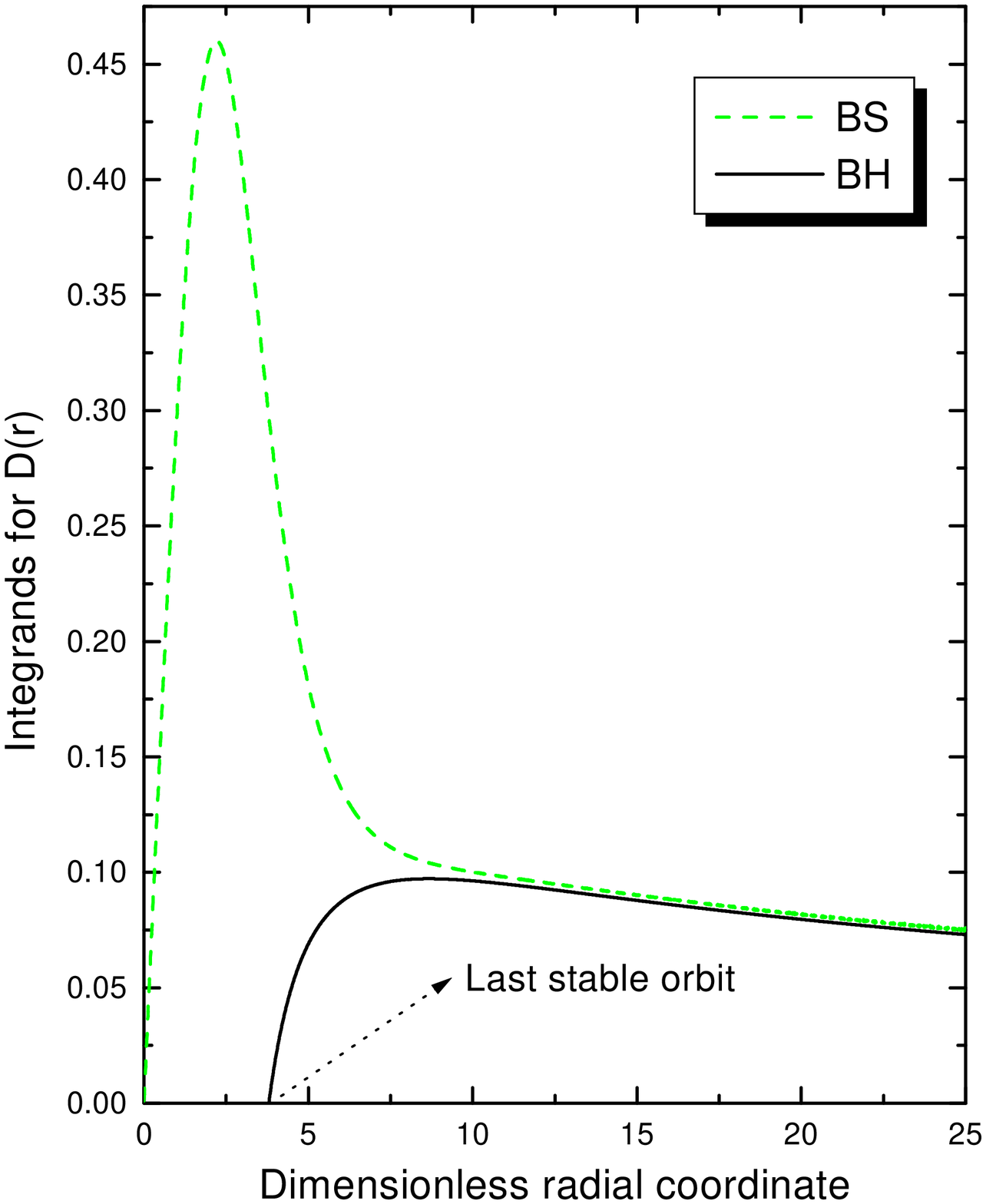}
\includegraphics[width=8.5cm,height=10cm]{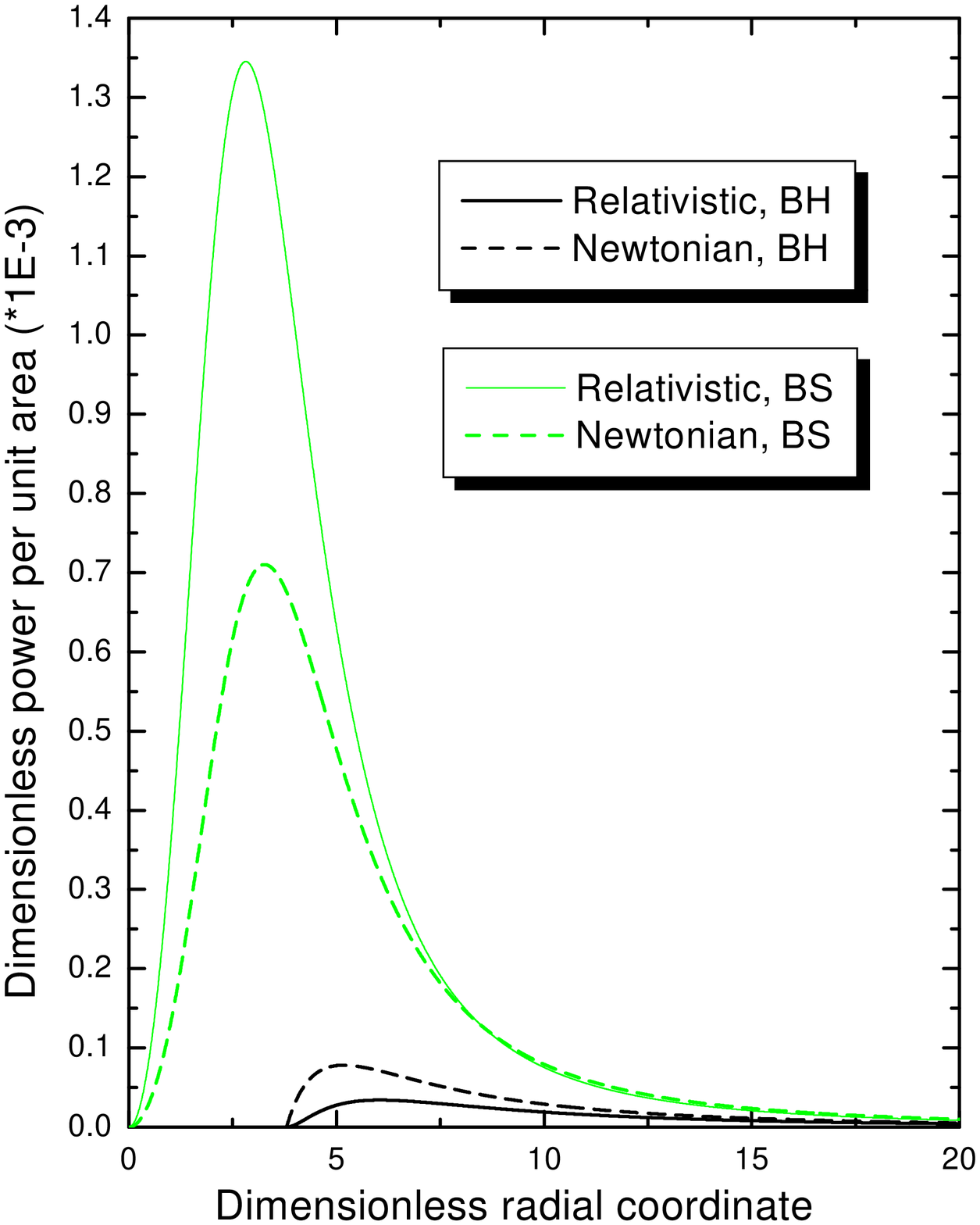}
\end{center}
\vspace{-1.5cm} \caption{Left: Integrands for D(r), both for a
black hole and a boson star of the same mass. The BH last stable
orbit is marked. Right: Relativistic power per unit area. In dash
lines, we show, for comparison, also the Newtonian results for
both cases.} \label{rel}
\end{figure}
\begin{figure}[t]
\vspace{-1.5cm}
\begin{center}
\includegraphics[width=8.5cm,height=10cm]{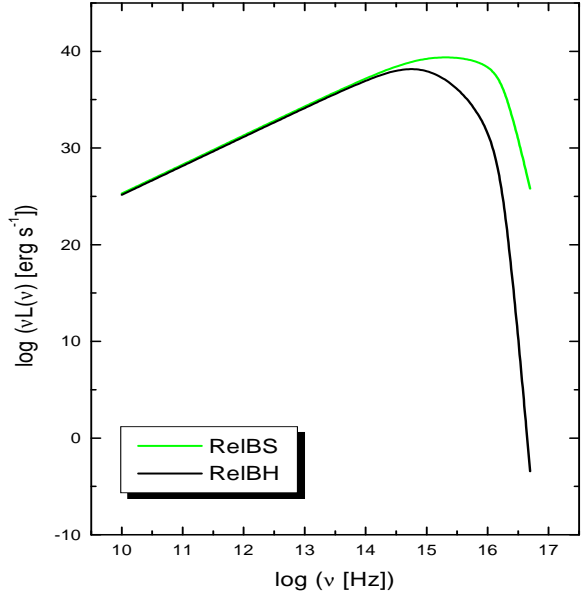}
\end{center}
\vspace{-1.5cm} \caption{Relativistic emission spectrum for an
accretion disc rotating around a static boson star. The black hole
case is shown for comparison. } \label{rel-spe}
\end{figure}

Figure \ref{rel} shows the results of some crucial intermediate
computation needed to get the relativistic spectrum, and, on the
right panel, the power per unit area in the relativistic disc. It
can be seen there that there is an interesting effect produced by
boson stars as central objects: While a relativistic treatment of
the disc reduces the expected emission from the innermost parts of
a disc rotating around a Schwarzschild black hole, it dramatically
enhances the expected one when a boson star is in the scenario.
This has the consequence of a hardening in the emission spectrum,
which is shown in Figure \ref{rel-spe}. The deviations shown
between the spectra are impressive. Table 3 shows some of the
values used to construct this latter figure.

\begin{table}[t]
\centering \caption{Comparison of relativistic spectra.}
\begin{tabular}{lll}
\hline\hline $\nu$ [1/s]  &   $\nu L(\nu)$ [erg s$^{-1}$] & $\nu
L(\nu)$ [erg s$^{-1}$] \\
 & BS & BH  \\ \hline
10$^{10}$ & $2.00 \times 10^{25}$ & $1.41  \times 10^{25} $ \\

10$^{12}$ & $2.00 \times 10^{31}$ & $1.41  \times 10^{31} $ \\

10$^{14}$ & $1.77 \times 10^{37}$ & $1.18  \times 10^{37} $ \\

10$^{16}$ & $1.36 \times 10^{39}$ & $3.46  \times 10^{33} $ \\
 \hline \hline
\end{tabular}
\end{table}

\section{Conclusions}

In this paper we have modeled a very simple accretion disc
rotating around a static supermassive boson star, although we have
given the scaling property that shows how to extend these results
to other mass domains (equivalently, to other single boson mass
cases). The disc was assumed steady, with a constant accretion
rate, and thin, so that the standard theory can be applied.
Throughout the paper, we have made a comparison of all results
with those obtained for discs rotating around Schwarzschild black
holes of the same mass. Our aim was to see whether the emissivity
properties of the accretion disc are noticeably changed when the
central object is. More complicated models for the accretion
process as well as more realistic models for the star (as those in
which the star is rotating) can be considered. We hope this work
will encourage further analysis. \\

\subsection*{Acknowledgments}
This work has been partially supported by CONICET and Fundaci\'on
Antorchas. The author is on leave from IAR, and acknowledges an
anonymous Referee for his remarks.

\end{document}